\documentclass[lettersize,journal]{IEEEtran}
\usepackage{amsmath,amsfonts}
\usepackage{algorithmic}
\usepackage{algorithm}
\usepackage{array}
\usepackage[caption=false,font=normalsize,labelfont=sf,textfont=sf]{subfig}
\usepackage{textcomp}
\usepackage{stfloats}
\usepackage{url}
\usepackage{verbatim}
\usepackage{graphicx}
\usepackage{cite}
\usepackage{color}
\usepackage{amsfonts,amssymb}
\usepackage{multirow}
\usepackage{makecell}
\usepackage{hyperref}
\hypersetup{
    colorlinks=true,
    linkcolor=blue,     
    urlcolor=blue,
    }
\begin{document}

\title{Federated Graph Neural Networks: Overview, Techniques and Challenges}

\author{Rui Liu,
        Pengwei Xing,
        Zichao Deng, 
        Anran Li,
        Cuntai Guan,~\IEEEmembership{Fellow,~IEEE,}
        Han Yu$^{*}$,~\IEEEmembership{Senior Member,~IEEE}
\thanks{The authors are with the School of Computer Science and Engineering, Nanyang Technological University, Singapore.

Han Yu is the corresponding author (Email: han.yu@ntu.edu.sg)}
\thanks{Manuscript received April 19, 2021; revised August 16, 2021.}}

\markboth{Journal of \LaTeX\ Class Files,~Vol.~14, No.~8, August~2021}%
{Shell \MakeLowercase{\textit{et al.}}: A Sample Article Using IEEEtran.cls for IEEE Journals}


\maketitle

\begin{abstract}
With its capability to deal with graph data, which is widely found in practical applications, graph neural networks (GNNs) have attracted significant research attention in recent years. As societies become increasingly concerned with the need for data privacy protection, GNNs face the need to adapt to this new normal. 
Besides, as clients in Federated Learning (FL) may have relationships, more powerful tools are required to utilize such implicit information to boost performance.
This has led to the rapid development of the emerging research field of federated graph neural networks (FedGNNs). This promising interdisciplinary field is highly challenging for interested researchers to grasp. The lack of an insightful survey on this topic further exacerbates the entry difficulty. In this paper, we bridge this gap by offering a comprehensive survey of this emerging field. We propose a 2-dimensional taxonomy of the FedGNNs literature: 1) the main taxonomy provides a clear perspective on the integration of GNNs and FL by analyzing how GNNs enhance FL training as well as how FL assists GNNs training, and 2) the auxiliary taxonomy provides a view on how FedGNNs deal with heterogeneity across FL clients.
Through discussions of key ideas, challenges, and limitations of existing works, we envision future research directions that can help build more robust, explainable, efficient, fair, inductive, and comprehensive FedGNNs.
\end{abstract}


\section{Introduction}
\IEEEPARstart{G}{raph} neural networks (GNNs) are powerful tools for dealing with graph-structured data \cite{wu2020comprehensive}. 
Graph-structured data are data samples connected by a graph topology. For example, molecular data are graph-structured data with atoms as the nodes and the bonds connecting them as the edges in the graph.
GNNs can improve the quality of node embedding by considering neighborhood information extracted from the underlying graph topology.
They have been widely adopted by diverse applications including drug discovery \cite{gilmer2017neural}, neuroscience \cite{ding2021lggnet}, social networks \cite{hamilton2017inductive}, knowledge graphs \cite{chen2020toward}, recommender systems \cite{ying2018graph} and traffic flow prediction \cite{cui2019traffic}. 

A well-trained GNN requires a large amount of training graph data, which may be distributed among multiple data owners in practice. 
Due to privacy concerns \cite{GDPR}, these data owners (a.k.a. clients) may not be willing to share the data, which leads to the problem of data isolation. 
Furthermore, the graph data stored by different clients are often non-independent and identically distributed (non-IID), which exacerbates the data isolation issue.
Such non-IID properties can manifest as differences in graph structures or node feature distributions across clients. 

Federated learning (FL), a distributed collaborative machine learning paradigm, is a promising approach to deal with the data isolation challenge \cite{li2020federatedsurvey,kairouz2021advances, li2021sample, li2021efficient}. 
It enables local models to benefit from each other, while keeping local data private \cite{Kairouz-et-al:2021, li2021privacy}.
In addition, the problem of learning personalized FL models in the presence of non-IID data has been extensively studied \cite{tan2022towards}. 
In FL, only model parameters or embedding features are shared among the participants without exposing potentially sensitive local data. This architectural design, combined with various cryptographic techniques, can provide effective protection of local data privacy. 
In some situations, FL participants have relationships with each other,  consisting of a graph topology with participants as nodes. This relationship graph may contain useful but implicit information, such as the participants' similarities and trust. 
Utilizing the topology information to boost FL performance remains a challenge. 

The confluence of these trends of development has inspired the emergence of the field of federated graph neural networks (FedGNNs) in recent years \cite{scardapane2020distributed}, which has witnessed rapid development in recent years.
Existing works such as \cite{zhang2021federated,he2021fedgraphnn} summarize FedGNNs into three categories: 1) FL clients containing multiple graphs, 2) FL clients containing sub-graphs, and 3) FL clients containing one node, according to the distribution of graph data. 
However, as technical research for these envisioned categories has not been extensively studied at the time, these early positioning papers only provided general ideas without specific works or problem descriptions for the different categories. Besides, there can be overlapping situations between Category 2 and Category 3. For example, in Category 2, if there are some edges connecting sub-graphs residing in different FL clients, these edges can be regarded as inter-client graphs, which also exist in Category 3. 
A recent survey \cite{fu2022federated} simplifies the three-category taxonomy with a two-category taxonomy based on the location of structural information: 1) structural information existing in the FL clients, and 2) structural information existing between FL clients. However, the problem-based sub-categories of the taxonomy are unreasonable as they only cover a limited subset of the FedGNNs literature. 

Currently, there is a lack of a comprehensive survey on FedGNNs that provides an insightful view on this critical topic for new researchers.
This paper bridges this important gap. The main contributions are as follows: 
\begin{itemize}
    \item We propose a two-dimensional (2D) taxonomy that categorizes existing works on FedGNNs from two perspectives: 1) the main taxonomy - how FL and GNNs are integrated together; and 2) the auxiliary taxonomy - how FedGNNs deal with heterogeneity across FL clients. We highlight the challenges, specific methods, and potential limitations for each category.
    \item We discuss commonly adopted public datasets and evaluation metrics in the existing literature for FedGNNs benchmarking, and offer suggestions on enhancing FedGNNs experiment design.
    \item We envision promising future directions of research towards building more robust, explainable, efficient, fair, inductive and comprehensive FedGNNs to enhance the trustworthiness of this field.
\end{itemize}


The rest of this paper is organized as follows. Key terminologies used in FedGNNs and the proposed 2D taxonomy are introduced in Section \ref{sec:Taxonomy}. The main challenges, techniques, and limitations of FedGNNs are reviewed in Section \ref{sec: GNN_assisted_FL}, \ref{sec:FL_assisted_GNN} for the main taxonomy and in Section \ref{sec:FL_aggregation_taxonomy} for the auxiliary taxonomy. Section \ref{sec:Datasets} summarizes applications, datasets, evaluation metrics and data partition methods of FedGNNs. Finally, we propose seven future directions towards building trustworthy FedGNNs in Section \ref{sec:Future_Direction}.

\section{Terminology and Taxonomy}
\label{sec:Taxonomy}

This section explains key terminologies in GNNs and FL, and introduces the proposed 2D FedGNNs taxonomy.

\subsection{Terminology}

GNNs are a class of deep learning models designed to perform feature embedding and inference on graph data.
They require two inputs: 1) a graph, which consists of \textit{nodes} and \textit{edges}, represented by an \textit{adjacency matrix} $\textbf{A}\in \mathbb{R}^{N\times N}$; and 2) their \textit{node features} $\textbf{X} \in \mathbb{R}^{N\times f}$. 
$N$ denotes the number of nodes and $f$ is the number of node features.
GNNs update the embedding of a given node by aggregating information from its neighboring nodes with the following function:
\begin{equation}
\textbf{X}^{(l+1)} = GNN(\textbf{A}, \textbf{X}^{(l)}, \textbf{W})
\label{eq:GNN}
\end{equation}
where $GNN(\cdot)$ indicates the graph aggregation function, which can be the mean, weighted average or max/min pooling methods. $\textbf{X}^{(l)}$ and $\textbf{X}^{(l+1)}$ represent the node embedding in the $l$-th and $(l+1)$-th layer. $\textbf{A}$ denotes the graph adjacency matrix. $\textbf{W}$ are the trainable model weights.

FL is a collaborative machine learning paradigm that trains a model across multiple data owners, without exchanging raw data. 
It has two main settings: \textit{horizontal} FL (HFL) and \textit{vertical} FL (VFL) \cite{FL:2019}. 
In HFL, the datasets in different clients have large overlaps in the feature space, but little overlap in the sample space.
In VFL, the clients have little overlap in the feature space, but large overlaps in the sample space.
In FL, data owners with sensitive local data can be referred to as \textit{clients} if they are coordinated by a central entity referred to as the \textit{server}. 
Under the HFL setting, based on the communication architecture, FL has two settings: \textit{centralized} FL and \textit{decentralized} FL. 
In centralized FL, the server coordinates the clients to jointly learn a model, while in decentralized FL, clients communicate with each other without a centralized server to jointly learn a model.


FL also involves an ``aggregation'' operation. 
Aggregation, in the context of FL, updates model parameters in the server with local model parameters uploaded by clients. For instance, it can be achieved with an averaging operation following FedAvg \cite{mcmahan2017communication}: 
\begin{equation}
L(\textbf{W}) = \sum_{c=1}^C\frac{N_c}{N_s} L_c(\textbf{W}) = \frac{1}{N_s}\sum_{c=1}^C\sum_{i=1}^{N_c}l_c(x_i,y_i,\textbf{w})
\label{eq:Fedavg}
\end{equation}
where $N_c$ denotes the number of samples in client $c$ and $N_s$ represents the total number of samples in all clients. $L(\textbf{W})$, $L_c(\textbf{W})$ and $l_c(\textbf{x}_i,y_i,\textbf{w})$ are global loss function, local loss function in client $c$ and local loss function on sample $\textbf{x}_i$ with model parameter $\textbf{w}$ and label $y_i$ in the client $c$.

To disambiguate between the aggregation operations in GNNs and FL, we refer to them as ``\textit{GNN aggregation}'' and ``\textit{FL aggregation}'', respectively in this paper. 

\begin{figure*}[t]
    \centering
    \includegraphics[width=2.05\columnwidth]{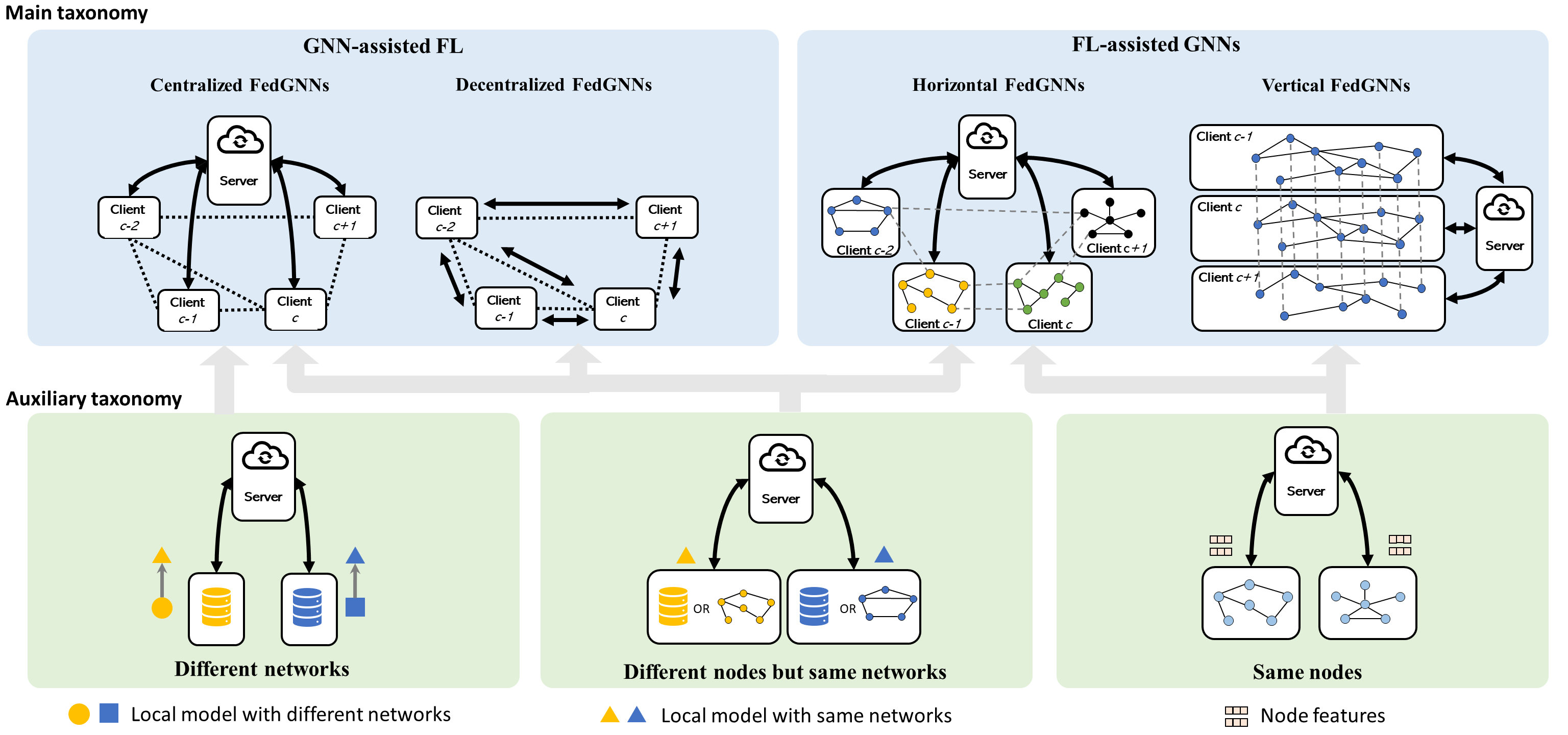}
    \caption{The proposed 2-dimensional taxonomy for FedGNNs research.}
    \label{fig:2D_taxonomy}
\end{figure*}

\subsection{The Proposed 2-Dimensional FedGNNs Taxonomy}


As shown in Figure \ref{fig:2D_taxonomy}, the proposed FedGNNs taxonomy consists of two dimensions. The first dimension focuses on the integration of FL and GNNs, which serves as the main taxonomy. The second dimension focuses on FL aggregation solutions dealing with different levels of graph data heterogeneity, which serves as the auxiliary taxonomy.
The main taxonomy can be divided into two categories:

\paragraph{GNN-Assisted FL} 
Approaches under this category focus on FL training with structured clients.  
Techniques used in GNNs are applied to assist FL model training.
In general, the existence of an inter-client graph structure (denoted by the dotted lines between clients in the GNN-assisted FL category in Figure \ref{fig:2D_taxonomy}), can be utilized by GNNs to improve the performance of existing FL algorithms. 
Works in this category can be further divided into two scenarios according to whether a central FL server exists or not: 1) centralized FedGNNs, and 2) decentralized FedGNNs. 
The central server is generally assumed to have a global view of the inter-client graph topology. It can leverage this view to: 1) train a GNN model to improve FL aggregation; or 2) help clients update their local models with a GNN model trained within the client. 
Without a central server, the inter-client graph topology must be given in advance by making each client hold a sub-graph so that clients can find their neighbors.

\paragraph{FL-Assisted GNNs}
Approaches under this category focus on training GNNs with isolated graph data silos.
FL algorithms are applied to assist GNN model training. 
In general, it is assumed that the graph data are stored distributedly and clients only have access to their local graph data.
FL is leveraged to train a global GNN model in this situation.
According to whether different clients share the same node IDs, FedGNNs under this category can be further divided into two scenarios: 1) horizontal FedGNNs, and 2) vertical FedGNNs.
In horizontal FedGNNs, clients have graph data consisting of nodes which are largely not overlapping. 
FL can help with the general case, but there might be missing edges between clients, which may require more complex solutions.
In vertical FedGNNs, clients share the same set of node IDs, but have different features. 
According to the different feature partitions, different solutions have been proposed.

The auxiliary taxonomy focuses on dealing with the heterogeneity among FL clients. They can be divided into three categories:
1) clients having nodes with the same IDs, 2) clients having different nodes but the same network structure, and 3) clients employing different network structures.
Different intermediate information is applied in FL aggregation for different categories.
For clients having the same nodes, node embedding features are uploaded to the FL server for aggregation. This can be found in vertical FedGNNs and some horizontal FedGNNs works with overlapping nodes.
For clients having different nodes but applying the same network structures for training, model weights and gradients are used for FL aggregation. This can be found in both scenarios of GNN-assisted FL and some horizontal FedGNNs works without overlapping nodes.
For the clients training local models with different network structures, the network structure can be modeled as a graph first and a GNN model is applied to it. Then, the GNN model weights or gradients can be used for FL aggregation. 
Currently, this can only be found in centralized FedGNNs works.


\section{GNN-assisted Federated Learning}
\label{sec: GNN_assisted_FL}

In this section, we review approaches under the GNN-assisted FL category in the main taxonomy. 
In some applications, clients have relationships with each other, which can be represented by a graph.
For example, road traffic monitoring sensors that are nearby each other tend to record similar traffic conditions.
An inter-client graph can be built from such graph-structured clients with each of them represented as a node in the graph. 
Due to the existence of the inter-client graph, GNNs algorithms are applied to assist the FL training process.
GNNs can leverage the graph in the FL system to address the non-IID problem across clients based on the assumption that clients who are closely related in the graph tend to share similar data distributions.
Besides, GNNs can also assist the FL system by modeling the neural network as a graph when local model architectures are different.


\begin{figure*}[t]
    \centering
\includegraphics[width=1.8\columnwidth]{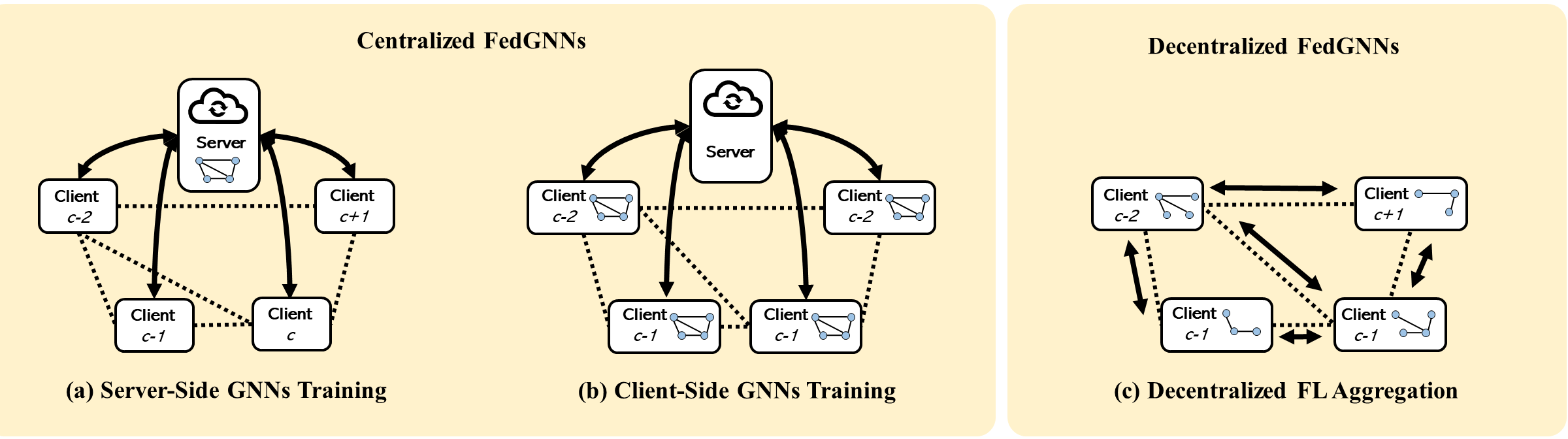}
    \caption{Illustrations of sub-scenarios in the centralized FedGNNs and decentralized FedGNNs (the dotted lines represent relationships among clients).}
    \label{fig:GNNsAssistFL}
    \vspace{-12pt}
\end{figure*}

\subsection{Centralized FedGNNs}

Centralized FedGNNs have a central server to coordinate clients (indicated by the arrow lines in Figure \ref{fig:GNNsAssistFL}) based on the inter-client graph.
Clients' local data do not necessarily need to be graph data. 
Depending on where the graph is stored, GNNs training can take place in the server (Figure \ref{fig:GNNsAssistFL}(a)), or the client (Figure \ref{fig:GNNsAssistFL}(b)).


\subsubsection{Server-Side GNNs Training}
A GNN model is trained in the server with the inter-client graph.
It assumes that neighboring clients tend to have similar local models or feature embedding.
The server first collects parameters from clients as it does in standard FL. The uploaded local model parameters are treated as the node features in the inter-client graph.
Then, it trains a GNN model with the uploaded parameters to facilitate FL aggregation. 
Finally, the updated parameters are sent back to the clients. 
The inter-client graph can be given in advance or extracted with a self-attention module during training \cite{wu2022multi}.
As the server has a separate GNN model, how to train both local models and the GNN model (Bi-level model) in the server simultaneously is a challenge.

\textbf{Bi-level Model Training:}
Works such as \cite{xingbig,chen2022personalized} design bi-level optimization schemes to train both local models and the GNN model with two types of objective functions: the local task objective functions $g_c(\cdot)$ for the local model training in the clients, and an objective function $f(\cdot)$ for GNN model training in the FL server:
\begin{equation}
\begin{aligned}
\min_{\phi} \quad & f(\phi,{\textbf{w}^*_c(\phi)|c\in{1,...,C}})\\
\textrm{s.t.} \quad  &\textbf{w}^*_c(\phi)\in \arg \min_{\textbf{w}_c} g_c(\phi,\textbf{w}_c)
\end{aligned}
\label{eq:BigFed}
\end{equation}
where $\phi$ denotes the trainable parameters in the GNN model and $\textbf{w}^*_c(\phi)$ is the local solution weight vector of client $c$.
Big-Fed\cite{xingbig} and SFL\cite{chen2022personalized} adopt different objective functions for $f(\cdot)$ to fulfill the assumption that neighboring clients' local models are similar.
Big-Fed proposed an unsupervised contrastive learning loss function, meanwhile, SFL\cite{chen2022personalized} proposed a supervised loss function with a graph smoothness regularization to train both local and global models.

Different from the bi-level optimization, some works \cite{li2022power,lee2022privacy} train the local models and the GNN model sequentially with separate objective functions.
For example, clients in \cite{lee2022privacy} train their local models with different local tasks. Then, the server trains a GNN model to fuse multi-task local estimates. By minimizing data reconstruction error with a graph regularization term, local estimates can be refined based on the clients' similarities.   
PDGNet \cite{li2022power} models the power allocation policy with a GNN model in the server to find the optimal power allocation policy. The objective function is to minimize the transmission error probability for all FL clients. The GNN model is trained with a primal-dual iterative approach.
CNNFGNN \cite{meng2021cross} and MLFGL\cite{wu2022multi} train the FL local models and the GNN model with only a local objective function. They perform alternating optimization to update clients' model weights with GNN model weights fixed and then update GNN model weights with the FL local model weights fixed, over multiple rounds.

\subsubsection{Client-Side GNNs Training}
\label{sssec:CentralizedFedGNNs_GNN_in_client}
GNN models are trained in the clients to solve two challenges: 1) data distribution heterogeneity, and 2) model heterogeneity.
The paradigm follows the general FL training procedure. A GNN model is trained in the client with model weights uploaded to the FL server. The server performs FL aggregation and distributes updated model weights to the clients for the next round of training. Meanwhile, building different graphs in the FL clients can solve different problems.

\textbf{Data Distribution Heterogeneity:}
Under this setting, it is assumed that each client has a global inter-client graph indicating the relationships among clients. 
Clients not only train the local models as they do in standard FL, but also train a GNN model with the global graph to obtain global knowledge from other clients to address the data distribution heterogeneity issue.  
FedCG \cite{caldarola2021cluster} builds a fully connected graph based on the similarity between clients' model weights or pattern features. A client trains a GNN with the graph to obtain the global embedding and then combines the local model embedding and the global embedding with a trainable weight.

\textbf{Model Heterogeneity:}
Standard FL is not well suited to deal with situations in which clients' local models are heterogeneous. 
HAFL-GHN\cite{litany2022federated} proposes a solution with the help of GNNs. It models the neural architecture in the client as a graph with each parametric layer as a vertex and the computational flows between layers as edges. The node features are initialized with a categorical (one-hot) feature indicating the layer type.
A GNN-based graph hyper network (GHN) processing the graph representation of architecture is trained to minimize the empirical risk of all clients. The output latent node features of GNNs are mapped back to the layer weights for training the original network.  
By converting a neural network into a graph and training it with a GHN model, heterogeneous model weights can be aggregated across FL clients indirectly by uploading local GHN weights to the FL server for aggregation.

\begin{table*}[t]
\centering
\caption{Summary of advantages and disadvantages for various GNN-Assisted FL scenarios.}
\resizebox*{1\linewidth}{!}{
\begin{tabular}{|m{0.1\textwidth}|m{0.1\textwidth}|m{0.2\textwidth}|m{0.4\textwidth}|}
\hline
 \textbf{Scenario} & \textbf{Sub-Scenario}  & \textbf{Advantages} & \textbf{Disadvantages}  \\ \hline
\multirow{2}{*}{\makecell[l]{Centralized \\FedGNNs}} & \makecell[l]{Server-Side \\GNNs Training} & 
\begin{itemize}
\item The server has more flexibility in FL aggregation to relieve non-IID problem.
\end{itemize} 
&
\begin{itemize}
\item Difficult to prove the convergence.
\item The server requires high computation costs when the inter-client graph is large.
\item Imprecise inter-client graph deteriorates performance.
\end{itemize} 
\\ \cline{2-4}
 & \makecell[l]{Client-Side \\GNNs Training}  
 & 
\begin{itemize}
\item Relieve the non-IID problem between clients.
\end{itemize} 
&  
\begin{itemize}
\item Shared inter-client graph may leak privacy.
\item Imprecise inter-client graph deteriorates performance. 
\end{itemize}  
\\ \hline
 \makecell[l]{Decentralized \\FedGNNs} & \makecell[l]{Decentralized \\ FL Aggregation}  &
\begin{itemize}
\item Relieve the non-IID problem with personalized local models in clients.
\item Do not require a central server.
\end{itemize} 
&
\begin{itemize}
\item Shared models may leak privacy between neighbors.
\item High communication cost between clients.
\item Clients with higher centrality are vulnerable. 
\item Need to re-train the model when new clients join.  
\end{itemize} 
\\ \hline
\end{tabular}
}
\label{tab:GNN_Assist_FL_Summary}
\end{table*}

\subsection{Decentralized FedGNNs}
\label{ssec:Decentalized_FedGNNs}

As illustrated in Figure \ref{fig:GNNsAssistFL}, decentralized FedGNNs do not have a central server to coordinate FL clients. 
Thus, performing decentralized FL model aggregation is a key challenge. 


\textbf{Decentralized FL Aggregation:}
Decentralized FL assumes that the clients can communicate with their neighbors. 
Existing works come up with two approaches to achieve decentralized FL aggregation in situations where clients are nodes related by a graph topology: 
1) updating the FL model parameters via weighted summation of model updates within the neighborhood, and 2)
updating the FL model parameters via graph regularization.

\textit{Weighted Summation of FL Model Parameters.}
In this approach, FL clients communicate with their neighbors and update their local models by aggregating their neighbors' local model parameters based on the graph topology connecting them:
\begin{equation}
\textbf{w}_c^{r+1} = \sum_{j\in \mathcal{N}(c)} a_{cj}\cdot \left[\textbf{w}_j^{r}\right]
    \label{eq:Weighted_summation}
\end{equation}
where $\textbf{w}_c^{r+1} \in \mathbb{R}^p$ denotes the local model parameters of client $c$ at round $r+1$, which can be a Bayesian model \cite{lalitha2019peer}, Gated Recurrent Units \cite{scardapane2020distributed} or GNNs \cite{pei2021decentralized}.
$[\cdot]$ is the encryption operation for data privacy protection, such as Diffie-Hellman key exchange \cite{pei2021decentralized} or secret sharing \cite{rizk2021graph}.
$a_{cj}$ is the [$c$-th row, $j$-th column] element in the adjacency matrix $\textbf{A}$ of the graph, which is assumed to reflect the local data distribution similarity between client $c$ and $j$.
$\mathcal{N}(c)$ is neighborhood of $c$ (including itself).  
All works under this section apply Eq. \eqref{eq:Weighted_summation} once per round (i.e., a client only aggregates models from its 1-hop neighbors). 

There are some variations of the above FL aggregation strategy with different foci.
To achieve faster convergence, DSGT \cite{Lu2020decentralized} utilizes the decentralized stochastic gradient tracking. 
To deal with a large-scale graph,
\cite{rizk2021graph} proposes a multi-server FedGNNs architecture
to increase the communication efficiency.
It assumes that there are multiple servers in the network related by a fixed graph topology and that there is no central server coordinating the network of servers. 
Clients under each server conduct FL model training following the classic centralized FL protocol. 
Once all the servers have aggregated their own clients' model updates, they perform inter-server model aggregation following Eq. \eqref{eq:Weighted_summation} among themselves.
PSO-GFML \cite{gogineni2022decentralized} enhances communication efficiency by only exchanging a portion of local model parameters with the servers. 
Instead of knowing the adjacency matrix in advance, the graph can be learned during training, similar to Graph Attention Networks (GATs) \cite{velickovic2018graph}. 
The edge weights of the inter-client graph are calculated based on the similarity between unlabeled graph embeddings \cite{tao2022semigraphfl} or hidden parameters \cite{yuan2022fedstn} in the corresponding clients.

\textit{Graph Regularization on FL Model Parameters.}
In this approach, graph Laplacian regularization is incorporated into the objective function  to make model parameters from neighboring clients similar in order to address the non-iid problem \cite{ortega2018graph}:
\begin{equation}
R(\textbf{W},\textbf{L})
= tr(\textbf{W}^T\textbf{L}\textbf{W})
= \frac{1}{2}\sum_{ij} a_{ij}\|\textbf{w}_i-\textbf{w}_j\|^2
\label{eq:Laplacian_regularization}
\end{equation}
where $\textbf{W} \in \mathbb{R}^{n\times p}$ denotes the model weights of neighboring clients. $\textbf{L} \in \mathbb{R}^{n\times n}$ is the Laplacian matrix of the graph topology between neighboring clients. $tr(\cdot)$ is the trace operation. $a_{ij}$ is the edge weight in the adjacency matrix connecting client $i$ and $j$. $\textbf{w}_i \in \mathbb{R}^p$ indicates the model parameters in client $i$.
Thus, for each client, the local objective function can be written as:
\begin{equation}
min_{\textbf{w}_c} L_c(\textbf{w}_c) + \lambda\cdot\frac{1}{2}\sum_{j\in \mathcal{N}(c)} a_{cj}\|\textbf{w}_j-\textbf{w}_c\|^2
\label{eq:Local_loss_Laplacian_regularization}
\end{equation}
where $\lambda$ indicates a balanceing weight and $\mathcal{N}(c)$ represents the neighbors of client $c$. Each client can only get the related information from their neigbors.

Multi-task learning can be performed with the above strategy. 
dFedU\cite{dinh2021new} assumes that each client has one task and a fully connected inter-client graph is given in advance. 
Once each client obtains the local updated models from its neighbors, it performs model updating with graph regularization. 
SpreadGNN \cite{he2021spreadgnn} assumes that each client solves multiple tasks. An inter-client task relationship graph is initialized from the task classifier model parameters.
Clients apply Decentralized Periodic Averaging SGD (DPA-SGD) to optimize the objective function and update model weights as well as their corresponding task relationship graph iteratively with a convergence guarantee.
Fed-ADMM \cite{wang2022heterogeneous} solves Eq. \eqref{eq:Local_loss_Laplacian_regularization} by proposing a decentralized stochastic version of the alternating direction method of multipliers (ADMM) algorithm with rigorous statistical guarantees of their estimators.



\subsection{Summary}
\label{ssec:GNN_Assist_FL_Summary}

In this section, we have discussed GNN-Assisted FL approaches, which leverage the GNN model training to improve FL aggregation. We now summarize the techniques in terms of their advantages and disadvantages, as listed in Table \ref{tab:GNN_Assist_FL_Summary}.

Centralized FedGNNs deal with the graph-structured FL system setting. 
With a GNN model trained in the FL server,  it has more flexibility in the FL model aggregation to deal with the non-IID problem among FL clients. The GNN model plays a trade-off between personalization on the client side and generalization on the server side. 
However, it is more difficult to prove the convergence with two objective functions (one for local models and the other for the global GNN model).
Besides, when it is applied to an FL system with a large-scale inter-client graph, the training cost of the GNN model in the server becomes enormous.
In addition, an imprecise inter-client graph in GNNs may deteriorate model performance.
With the inter-client graph stored by the clients with the GNN model trained locally, it can also relieve the non-IID problem. However, the locally stored inter-client graph may leak other clients' private information. Besides, it also faces the same issue as above when the inter-client graph is imprecise. 

Decentralized FedGNNs are designed for the serverless graph-structured FL setting.
It can address the non-IID problem among FL clients without a central server by making clients communicate with their neighbors directly. 
Due to the difference in the neighborhood for each client, the eventual aggregated model for each client is a personalized local model.
However, such peer-to-peer learning with model weights shared directly between neighboring clients may result in privacy leakage and incur high communication costs. 
Besides, clients with higher centralities are vulnerable to attacks. 
In addition, it is not continual learning as the local models need to be retrained when a new client joins as their neighbor.

\section{FL-Assisted Graph Neural Networks}
\label{sec:FL_assisted_GNN}
In this section, we introduce FL-Assisted GNNs in the main taxonomy.
In some applications, graph data are isolated and stored by different clients.
How to train GNN models with isolated graph data, while protecting data privacy, becomes the key challenge. 
Moreover, isolated graph data following different data distributions may lead to the non-IID problem, which is another major research problem. 
Due to the isolation of graph data, FL algorithms are applied to assist the GNN model training process. 
FL, as an emerging tool to process distributed data with privacy protection, can help train GNN models with distributed graph data. Moreover, personalized FL\cite{tan2022towards} can relieve the non-IID problem among clients.


\begin{figure*}[t!]
    \centering
\includegraphics[width=1\linewidth]{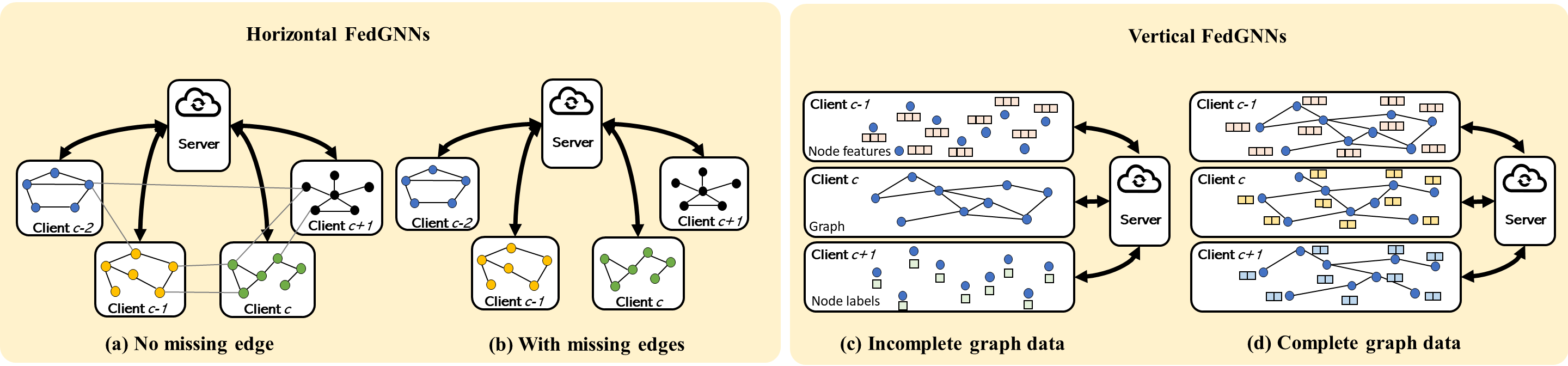}
    \caption{Illustrations of the sub-scenarios of horizontal FedGNNs and vertical FedGNNs.}
    \label{fig:FLAssistGNNs}
    \vspace{-12pt}
\end{figure*}

\subsection{Horizontal FedGNNs}
\label{ssec:Horizontal_FedGNN}


Horizontal FedGNNs refer to the situation whereby clients share the same node feature spaces but different node IDs. 
Each client has at least one graph or a set of graphs. 
There can be nodes stored in different clients which are connected by edges.
There are two sub-scenarios in horizontal FedGNNs.
The first assumes that these edges are retained by clients or no such edges exist between clients, referred to as ``no missing edge'' (Figure \ref{fig:FLAssistGNNs}(a)).
The second assumes that such edges connecting nodes in different clients exist but are missing, referred to as ``with missing edges'' (Figure \ref{fig:FLAssistGNNs}(b)).
Note that the works in this section focus on FL clients having graphs containing local information of neighboring clients, whereas the works in Section \ref{sssec:CentralizedFedGNNs_GNN_in_client} focus on FL clients knowing the global FL network topology. 



\subsubsection{No Missing Edge}
\label{sssec:HFedGNN_NoMissingEdge}

The common strategy for horizontal FedGNNs without missing edges is to train local GNN models in the clients to learn the local graph representations or node embedding first. Then, an FL algorithm is applied on top of it. The FL server collects the model parameters or gradients from clients for FL aggregation as described in Eq. \eqref{eq:Fedavg} \cite{du2022malicious, peng2022domain}, and sends back the updated parameters to the clients for the next round of training.
FedGNNs works under this setting are introduced according to the research problems they solve. 
GNNs-related research problems focus on non-IID problems in graph data, distributed spatial-temporal or large-scale graph data embedding, and distributed neural architecture search problems. 
FL-related research problems focus on the improvement of FL aggregation and privacy protection.

\textbf{Non-IID Problem in Graph Data:} Graph data in the clients from different domains follow heterogeneous data distributions. How to train a global GNN model with non-IID graph data becomes an important problem in this setting. 
Some global model personalized FL solutions\cite{tan2022towards} are borrowed to solve it: model-based approaches and data-based approaches.

Model-based approaches improve the adaptation performance of the local model or learn a powerful global FL model for the future personalization of each client, which includes model interpolation, regularized local loss, meta-learning, etc. 
ASFGNN\cite{zheng2021asfgnn} and FedEgo \cite{zhang2022fedego} apply the model interpolation technique in the client. The final model for the client is a combination of the global model and the local model.  The percent of the local model in updating process is controlled by a mixing weight, which can be the Jensen-Shannon divergence\cite{zheng2021asfgnn} or the earth mover distance (EMD)\cite{zhang2022fedego} between local and global data distribution.
FedAlign\cite{lin2020improving} adds an optimal transport (OT) distance-based regularization term between local and global models in the loss function to minimize the model divergence as FedProx \cite{li2020federated}. 
GraphFL \cite{wang2020graphfl} applies a meta-learning training scheme to mitigate the non-IID problem. Inspired by the model-agnostic meta-learning (MAML) \cite{finn2017model}, it finds a good initial model that can be fast adapted to clients after a few local updates. 

Data-based approaches target to decrease the statistical heterogeneity of client data distributions using sample re-weighting, clustering, manifold learning, etc.
FLIT\cite{zhu2021federated} solves the non-IID data problem by re-weighting samples based on their prediction confidence. To make the local training more consistent across clients and avoid overfitting the local data, 
it put more weight on the samples the local model is less confident in the prediction results than the global model.
GCFL+ \cite{xie2021federated} solves the non-IID problem by clustering clients based on the gradients of the Graph Isomorphism Networks (GIN) model \cite{xu2018powerful} from each client. General FL aggregation is applied within each cluster. 
FMTGL \cite{liu2022federated} relieves the non-IID problem by obtaining universal task representation with a fusion module shared across clients. 
The key to getting the universal task representation is to process the multi-source representation matrices on a common fusion space, consisting of several learnable support vectors. 

\textbf{Distributed Spatial-Temporal Graph Data Embedding:} 
As the local graph topology changes over time or each node contains time-series data,  it is necessary to embed the spatial-temporal information for the distributed graph data.

STFL\cite{lou2021stfl}, Feddy\cite{jiang2022federated} and 4D-FED-GNN+ \cite{gurler2022federated} deal with spatial-temporal graph data embedding differently. STFL\cite{lou2021stfl} doesn't consider the temporal relationship in the graph embedding. It treats the graph data at each time step as one sample and predicts the label for each of them. Feddy  \cite{jiang2022federated} considers temporal information in the graph embedding by applying dynamic GNNs. 
4D-FED-GNN+ \cite{gurler2022federated} focuses on the evolution graph learning task with missing time points. Each client trains a GNN model for each time step.
GNN acts as a generator or a self-encoder based on the data availability at the current and next time step. It improves the predictive performance of local models, while benefiting from other clients with data corresponding to the missing time points.

\textbf{Distributed Large-Scale Graph Data Embedding:} 
High computation costs hinder GNNs training on large-scale graphs.
To reduce the computation cost of GNN model training, FedGraph \cite{chen2021fedgraph} design a sampling policy for the server based on reinforcement learning. In each round, the server refines the sampling strategies (i.e., the number of nodes to be sampled) and the GNN model parameters for the clients.

\textbf{Distributed Neural Architecture Search:}
Designing a suitable architecture for the local GNN model is beneficial for the global model's performance. 
Existing works pay attention to the GNNs architecture search problem. HAFL-GHN \cite{wang2021fl} proposes a federated evolutionary optimization strategy to search for suitable GCN architectures. It applies a GCN SuperNet and a weight-sharing strategy to reduce the searching time so that the proposed algorithm can find better GCN models in a short time.
The approach in \cite{balik2022investigating} searches for the GNN models that can produce the most reproducible features.  
The model is selected based on a reproducibility matrix between paired GNN models. The reproducibility value between the two GNNs is defined by the overlap ratio of the top $K$ reproducible features of them, which are selected according to the last layer of the models.

\textbf{FL Aggregation Improvement:} 
The general FL aggregation approach is described as follows:
\begin{equation}
    \textbf{w}^{r+1}=\sum_{c=1}^C\eta_c\cdot \textbf{w}_c^r
    \label{eq:FL_aggregation}
\end{equation}
where $\textbf{w}^r$ and $\textbf{w}^{r+1}$ denotes the model weights in $c$-th client at the round $r$ and the global model weights in the server for the next round. $\eta_c$ represents the importance of client $c$ in FL aggregation. 
However, the definition of $\eta_c$ not only depends on the number of samples in the clients, it also affects FL model aggregation performance.

Some works improve the FL aggregation by defining $\eta_c$ based on given factors.
Fed-CBT\cite{bayram2021federated} defines it with the training round index. It puts more weights on the clients with latest updates.
In \cite{hu2022fedgcn}, the weight is defined by a trainable attention mechanism based on the global model parameters and local model parameters layer-wisely. 

4D-FED-GNN+ \cite{gurler2022federated} improves the FL aggregation with a mixed federation strategy. The server alternates two FL aggregation methods: FedAvg and model exchange methods \cite{matsuda2022fedme,mao2020fedexg}, to address the non-IID problem.
CTFL \cite{zhang2022graph} improves the communication efficiency of FL aggregation as the number of parameters in local GNNs is enormous. CTFL clusters clients based on the closeness of their local model parameters with a divide-and-conquer strategy. Only one representative local model of each cluster is sent to the server for FL aggregation to reduce communication costs.

\textbf{Privacy Protection:} 
By sharing model parameters and graph topology, FedGNNs have large attack surfaces.
Some works pay more attention to privacy protection. FeSoG \cite{liu2021federated} uploads encrypted gradients using dynamic local differential privacy (LDP) to the server for FL aggregation.  Clients in FedGraph \cite{chen2021fedgraph} encrypt local training with Trusted Execution Environment (TEE) and the server encrypts global model aggregation with secure multi-party computation (MPC) or homomorphic encryption (HE).
FedEgo \cite{zhang2022fedego} protects the graph privacy by constructing mashed ego-graphs in the client. The global structure is protected by sampling neighboring nodes with a fixed size for the central node to construct the ego-graphs in the clients.
Local graph embedding is anonymous by averaging a batch of ego-graphs (mixup or mashed) before being uploaded to the server.


\subsubsection{With Missing Edges}
\label{sssec:HFedGNN_MissingEdge}
In this scenario, it assumes that some edges connecting nodes in different clients are missing.
The missing edges can be divided into two types: 
1) edges between nodes with different node IDs, and 2) edges between aligned nodes in different clients.
For the first type, the common strategy for it is to amend the local graph by reconstructing the missing edges as a complete local graph can ensure a high quality graph representation and edges between clients can mitigate the non-IID data problem to some extend.
For the second type, knowledge graph (KG) completion is an important application under this setting. The key strategy is to transform information between aligned nodes across clients to help local KG embedding completion.
Once the local graphs are amended, the FL algorithm is applied to assist GNNs training in the same way as the ``No missing edge'' works. 
In this section, we summarize the existing works for these two sub-scenarios.


\textbf{Local Graph Amendment:}
Edge generators or node generators, located in the server, clients or a third party, are designed to amend the local graphs. 
Once the local graphs are complete, GNN models are trained on them in the clients and a FL aggregation is applied in the server to obtain a global model. 

FASTGNN \cite{zhang2021fastgnn} proposes a simple edge generator in the server. It reconstructs the missing edges between clients with Gaussian randomly generated edges and broadcasts them to all clients to update their local graphs.
The edge generator in FedGL \cite{chen2021fedgl}, located in the server, can generate a global pseudo graph with node embeddings uploaded by FL clients, and distributes it to the clients to amend their local graphs for GNN model training. 

Instead of generating missing edges directly, some works \cite{zhang2021subgraph,peng2022fedni} designed generative models to recover missing neighborhood node embedding first and then reconstruct missing edges based on them.
FedSage+ \cite{zhang2021subgraph} proposes a node feature generator in the client. 
To train the generator, the client holds out some existing edges randomly in the local graphs. The generator, equipped with a Gaussian noise generator, is trained to predict the number of missing neighborhood nodes and reconstruct hold-out neighborhood node features.
Once the sub-graph is updated, each client trains a GNN model continually, GraphSage \cite{hamilton2017inductive}, and uploads the model parameters to the server for FL aggregation.
FedNI \cite{peng2022fedni} improves the node generator by adding a discriminator to identify if the node features come from the generator or the real missing neighbor. This Generative Adversarial Network (GAN)-based generator \cite{goodfellow2020generative} improves the quality of generated node features. Besides, it removes nodes in the client with a breadth-first search, instead of random selection. 

DP-FedRec \cite{qiu2022privacy} leverages a Private Set Intersection (PSI) to extend the local graph and relieve the non-IID problem. Client $i$ and $j$ execute PSI protocol to get the intersected vertex. The client $i$ extends the edges and vertex for the intersected vertex within k-hop from the client $j$.
To protect privacy, all the clients add noise to the graph data (apply DP) before the local graph extension step.

FedPerGNN \cite{wu2022federated} and FedGNN \cite{wu2021fedgnn} amend their local graphs in a more safe way. They introduce a third-party server, which only deals with graph expansion for the clients. The original central server first generates and sends a public key to clients for local node IDs and embedding encryption. Then clients upload the ciphertexts to the third-party server. The third-party server locates the interacted nodes by checking the ciphertexts of their IDs and distributes encrypted node embedding to the clients to amend their local graphs for the following local GNNs training. 

Some works \cite{du2022federated,yao2022fedgcn} pay more attention to communication efficiency with simple graph amending methods. In \cite{du2022federated}, the client sends a request via the central server to the corresponding clients to get the missing embedding by neighbor sampling. It proposes an algorithm that can find an optimal sampling interval that achieves the best trade-off between convergence and running time. FedGCN \cite{yao2022fedgcn} allows clients to collect 1-hop or 2-hop averaged neighbor node features from other clients once at the beginning of the training to amend missing information. Then each client trains the local GNN model and the server collects local model weights for FedAvg-based FL aggregation. FedGCN also provides a theoretical analysis on the trade-off between the convergence rate and communication cost under different data distributions.


\textbf{Knowledge Graph Completion:}
KGs from different domains may contain the same entities. 
How to improve KG embedding quality with the help of other KGs without leaking privacy is a challenge. The key part is to transform information between aligned embedding across clients.  

FKGE \cite{peng2021differentially} designs a revised GAN-based module \cite{jordon2018pate} to translate the aligned entity and relation embedding between paired KGs.
If the paired KGs are improved, the refined embedding is broadcast to other KGs.
FedE \cite{chen2021fede} designs an overall entity table in the server to record all unique entities from clients. 
The server applies FedAvg FL aggregation on the aligned entity embedding in the table. Once finished, the updated entity embeddings are distributed to clients. The clients update entity embedding based on KG embedding methods with a self-training contrastive learning loss.

To tackle the privacy leakage issue in FedE, clients in FedR \cite{zhang2022efficient} upload relation embeddings instead of entity embeddings since the server cannot infer entity embedding given only relation embedding. 
To further protect the privacy, Secure Aggregation \cite{bonawitz2017practical} is applied to the relation embedding before being uploaded to the server and the relation table in the server is obtained via Private Set Union (PSU).

FedEC \cite{chen2022federated} improves FedE with the non-IID problem by adding a regularization term in the loss. It can increase the similarity between global and local entity embedding in the current round and decrease the similarity between local entity embedding in the current and the last round.

\begin{table*}[t]
\centering
\caption{Summary of advantages and disadvantages for different scenarios of FL-assisted GNNs.}
\resizebox*{1\linewidth}{!}{
\begin{tabular}{|m{0.1\textwidth}|m{0.1\textwidth}|m{0.2\textwidth}|m{0.4\textwidth}|}
\hline
 \textbf{Scenario} & \textbf{Sub-Scenario}  & \textbf{Advantages} & \textbf{Disadvantages}  \\ \hline
\multirow{2}{*}{\makecell[l]{Horizontal \\FedGNNs}} & \makecell[l]{No missing \\edge} & 
\begin{itemize}
\item Train GNN models with isolated graph data.
\item Relieve the data heterogeneity across clients.
\end{itemize} 
&
\begin{itemize}
\item Biased graphs may cause unfairness.
\item Few works encrypt the model weights.
\item Vulnerable to malicious attacks.
\item High communication costs.
\item The framework only works with limited GNN models. 
\end{itemize} 
\\ \cline{2-4}
 & \makecell[l]{With missing \\edges}  
 & 
\begin{itemize}
\item Recover missing information between clients.
\item Others are the same as above.
\end{itemize} 
&  
\begin{itemize}
\item Imprecise amended local graph deteriorates performance.
\item Shared node feature may leak privacy when amending local graph.
\item Others are the same as above.
\end{itemize}  
\\ \hline
 \multirow{2}{*}{\makecell[l]{Vertical \\FedGNNs}} & \makecell[l]{Incomplete \\ graph data}  &
\begin{itemize}
\item Train a GNN model with isolated graph data.
\end{itemize} 
&
\begin{itemize}
\item Number of clients is limited.
\item Vulnerable to malicious attacks.
\item The framework only works with limited GNN models.
\end{itemize} 
\\ \cline{2-4}
 & \makecell[l]{Complete \\graph data}  
 & 
\begin{itemize}
\item Train a GNN model with isolated graph data.
\end{itemize} 
&  
\begin{itemize}
\item Vulnerable to malicious attacks.
\item The framework only works with limited GNN models.
\end{itemize} 
\\ \hline
\end{tabular}
}
\label{tab:FL_Assist_GNN_Summary}
\end{table*}

\subsection{Vertical FedGNNs}
\label{ssec:VFedGNN}



Vertical FedGNNs assume that clients hold nodes with completely overlapping node IDs but different feature spaces.
Clients train a global GNN model with features from different clients with the help of FL.
There are two sub-scenarios in vertical FedGNNs.
The first assumes that graph topology, node features and node labels are owned by different clients. That is, clients do not have complete graph data (neither node features nor graph topology) (Figure \ref{fig:FLAssistGNNs}(c)).
The second assumes that only node feature spaces are owned by different clients. The graph topology is available for all clients (Figure \ref{fig:FLAssistGNNs}(d)).

\subsubsection{Clients with Incomplete Graph Data}
In this setting, different clients contain part of graph data. With three clients in the system, one client owns the node features, one owns the graph topology and one owns the node labels. 
Or one gets node features, and the other owns the rest if there are only two clients in the system.
How to make these clients work together, while protecting their privacy is a key challenge.

Instead of sharing the original adjacency matrix of the graph, SGNN \cite{mei2019sgnn} calculates a Dynamic Time Warping (DTW) algorithm-based similarity matrix to convey the same graph topology but conceal the original structure.
To protect the privacy of the node features, one-hot encoding is applied to map the original features to a matrix. 
Then the information from different clients is uploaded to the server to train a global GNN model for a node classification task.

FedSGC \cite{cheung2021fedsgc} assumes that there are only two clients without a central server. Graph topology and node features are owned by two clients. The client who has the node labels is the active party to create encryption key pairs. Clients encrypt the sensitive information using additively homomorphic encryption (AHE) before sending them to the other party for the GNN model parameter updating.

\subsubsection{Clients with Complete Graph Data}
In this setting, clients contain complete graph data including graph topology and node features. But their node feature types are different. Concatenating the node features is the key strategy.

FedVGCN \cite{ni2021vertical} assumes that there are only two clients with a central server.  
For each iteration, the two clients transfer intermediate results to each other under homomorphic encryption. 
The server is in charge of creating encryption key pairs for clients and doing FL aggregation for the model. 
Clients in VFGNN \cite{ijcai2022p272} first encrypt the node embedding with differential privacy (DP) and then integrate node features in a semi-honest server via mean, concatenation, or regression calculations as FL aggregation. 
Once training is complete, the client who owns node labels receives the updated node embedding from the server to perform node prediction.
FML-ST \cite{li2022federated} assumes that there is a global pattern graph shared by all clients with the same nodes. They fuse the local spatial-temporal (ST) pattern and global ST pattern using a multi-layer perceptron (MLP) with concatenated patterns as inputs. Clients leverage the global pattern to personalize their local pattern graph by evaluating the difference between global and local pattern graphs. 
Graph-Fraudster \cite{chen2022graph} studies the adversarial attacks on the local raw data and node embedding. It proves that differential privacy (DP) mechanism and top-k mechanism are two possible defenses to the attacks.

\subsection{Summary}
\label{ssec:FL_Assist_GNN_Summary}

In this section, we have discussed FL-assisted GNNs approaches, which leverage FL to assist GNN model training in a distributed setting. We now summarize the techniques in terms of their advantages and disadvantages.

Horizontal FedGNNs deal with clients having graph data with different node IDs. 
They can train GNN models with isolated graph data and relieve the data heterogeneity problem across clients. 
However, they generally overlook the heterogeneity in the graph topology. Biased graphs may cause unfairness in FedGNNs.
Besides, the privacy protection capability achieved by existing works is generally low. For example, few works consider encrypting the model weights before sending them to the server, making current works lack robustness against malicious attacks.
In addition, communication costs are high when the local graph model size or the number of clients is large.
Finally, existing approaches only work with limited basic GNN models. More advanced GNN models need to be included. 
Approaches capable of dealing with more difficult situations where edges between clients are missing can recover some missing information to improve performance. However, if the amended local graph is imprecise, the model performance may deteriorate. Besides, some approaches require clients to share some node features with the neighbors to repair local graphs, which can cause privacy leakage. 

Vertical FedGNNs deal with clients having graph data of different node features but the same node IDs.
They can help clients train a GNN model with isolated graph data. 
However, for the incomplete graph data setting, existing approaches only support two clients (pair-wise federated learning) or three clients, which is insufficient.
Besides, these systems are vulnerable to malicious attacks and they only work with limited basic GNN models. 

\begin{table*}[ht!]
\centering
\caption{Summary of FedGNNs literature in the 2D taxonomy.}
\resizebox*{1\linewidth}{!}{
\begin{tabular}{|lll|l|l|l|}
\hline
\multicolumn{3}{|l|}{} & \textbf{Same Nodes} & \textbf{Different Nodes, Same Network} & \textbf{Different Networks} \\ 
 \hline
\multicolumn{1}{|l|}{\multirow{4}{*}{\makecell[c]{ \textbf{GNN-} \\ \textbf{Assisted} \\ \textbf{FL}}}} & \multicolumn{1}{l|}{\multirow{2}{*}{\makecell[c]{Centralized\\ FedGNNs}}} & \makecell[l]{Server-side GNN\\ training} & nil & \makecell[l]{CNFGNN\cite{meng2021cross}, MLFGL\cite{wu2022multi},\\Big-Fed\cite{xingbig}, SFL\cite{chen2022personalized}, PDGNet \cite{li2022power}, \cite{lee2022privacy}} & nil \\ \cline{3-6} 
\multicolumn{1}{|l|}{} & \multicolumn{1}{l|}{} & \makecell[l]{Client-side GNN \\training} & nil & FedCG\cite{caldarola2021cluster} & HAFL-GHN\cite{litany2022federated} \\ \cline{2-6} 
\multicolumn{1}{|l|}{} & \multicolumn{1}{l|}{\makecell[c]{Decentralized\\FedGNNs}} & \makecell[l]{Decentralized FL\\ aggregation} & nil & \makecell[l]{D-FedGNN\cite{pei2021decentralized}, DSGT \cite{Lu2020decentralized}, \cite{lalitha2019peer}, \cite{scardapane2020distributed}, \cite{rizk2021graph},\\PSO-GFML\cite{gogineni2022decentralized}, FedSTN\cite{yuan2022fedstn}, SemiGraphFL\cite{tao2022semigraphfl},\\
Fed-ADMM\cite{wang2022heterogeneous}, dFedU\cite{dinh2021new}, SpreadGNN\cite{he2021spreadgnn}} & nil \\
\hline
\multicolumn{1}{|l|}{\multirow{4}{*}{\makecell[c]{ \textbf{FL-} \\ \textbf{Assisted}\\ \textbf{GNNs}}}} & \multicolumn{1}{l|}{\multirow{2}{*}{\makecell[c]{Horizontal\\ FedGNNs}}} & No   missing edge & nil & \makecell[l]{Fed-CBT\cite{bayram2021federated}, STFL\cite{lou2021stfl}, \cite{balik2022investigating} \\4D-FED-GNN+ \cite{gurler2022federated}, FedGCN\cite{hu2022fedgcn},\\FedGraph\cite{chen2021fedgraph}, Feddy\cite{jiang2022federated}, FeSoG\cite{liu2021federated},\\FL-AGCNS\cite{wang2021fl}, ASFGNN\cite{zheng2021asfgnn},\\FLIT/FLIT+\cite{zhu2021federated}, GCFL/GCFL+\cite{xie2021federated}, \\Fed-RGCN\cite{lin2020improving}, GraphFL\cite{wang2020graphfl}, FedEgo\cite{zhang2022fedego}\\ GraphSniffer \cite{du2022malicious}, DA-MRG\cite{peng2022domain}, \\ CTFL\cite{zhang2022graph}, FMTGL \cite{liu2022federated}} & nil \\ \cline{3-6} 
\multicolumn{1}{|l|}{} & \multicolumn{1}{l|}{} & With missing edges & \makecell[l]{FedE\cite{chen2021fede}, FedR\cite{zhang2022efficient},\\ FedEC\cite{chen2022federated}, FKGE\cite{peng2021differentially},\\FedGL\cite{chen2021fedgl}} & \makecell[l]{FedGL\cite{chen2021fedgl}, FASTGNN\cite{zhang2021fastgnn}, FedNI\cite{peng2022fedni},\\ FedSage/FedSage+\cite{zhang2021subgraph}, FedPerGNN\cite{wu2022federated},\\FedGNN\cite{wu2021fedgnn}, DP-FedRec\cite{qiu2022privacy}, \cite{du2022federated},  FedGCN\cite{yao2022fedgcn}} & nil  \\ \cline{2-6} 
\multicolumn{1}{|l|}{} & \multicolumn{1}{l|}{\multirow{2}{*}{\makecell[c]{Vertical \\FedGNNs}}} & Incomplete graph data & 
FedSGC\cite{cheung2021fedsgc}, SGNN\cite{mei2019sgnn} &  nil & nil \\ \cline{3-6} 
\multicolumn{1}{|l|}{} & \multicolumn{1}{l|}{} & Complete graph data & \makecell[l]{Graph-Fraudster\cite{chen2022graph},\\FedVGCN\cite{ni2021vertical},\\VFGNN\cite{ijcai2022p272}, FML-ST\cite{li2022federated}}  & nil & nil \\ \hline
\end{tabular}
}
\label{tab:2D_taxonomy_works}
\end{table*}

\section{The Auxiliary Taxonomy}
\label{sec:FL_aggregation_taxonomy}
In this section, we discuss the proposed auxiliary taxonomy for FedGNNs. 
According to the level of heterogeneity of local data and models across FL clients, existing works can be divided into three categories (with increasing heterogeneity levels): 1) clients with the same nodes, 2) clients with different nodes but the same network structure, and 3) clients with different network structures.
To implement FL aggregation under different situations, diverse strategies have been proposed with various intermediate information exchanged among FL clients (Table \ref{tab:2D_taxonomy_works}).

\subsection{FL Clients with the Same Nodes}
In this category, it is assumed that clients have the nodes with the same set of nodes but different types of node features. Vertical FedGNNs and part of horizontal FedGNNs with overlapping nodes (e.g., knowledge graph completion tasks) belong to this category. 
In this situation, FL clients usually upload the node feature embeddings to the FL server, where different feature embeddings of the same node are aggregated for knowledge transfer among clients. Besides, model weights can also be shared among clients to jointly train a global FL model in vertical FedGNNs.

\subsection{FL Clients with Different Nodes but the Same Network Structure}
In this category, it is assumed that FL clients have different nodes, but with the same network architecture.  
Most GNN-Assisted FL works and horizontal FedGNNs works belong to this category as their clients contain different nodes or samples.
Since the node IDs are different under this setting, node feature embedding cannot be uploaded to the server for FL aggregation. 
According to GNN aggregation:  $\textbf{X}^{(l+1)}=\textbf{A}\textbf{X}^{(l)}\textbf{W}$, the size of the trainable matrix $\textbf{W}$ is not related to the graph topology, but related to the dimensions of node features and output features.
Thus, the embedding of the entire graph \cite{yuan2022fedstn,li2022federated}, trainable model weights, and gradients in the GNN model can be uploaded to the server for FL aggregation. 
Whether model weights or gradients are to be uploaded depends on the preference between training speed and model performance \cite{mcmahan2017communication}.
Uploading model weights to the server allows each client to perform multiple epochs of local training. However, its local model update direction may deviate from the global FL model. 
Uploading gradients allows the client to closely follow the latest global optimization direction in every updating step. However, the frequent communication between clients and the server may increase communication costs during the training process.

\subsection{FL Clients with Different Network Structures}
Works in this category deal with a more difficult situation that network architectures in clients are different. 
GNN model weights can have the same size as long as the number of the node features is the same across clients. However, in this situation, the number of node features is different.
Currently, only \cite{litany2022federated} provides a solution. It converts the local network architecture into a graph with each layer as a vertex and the layer type as the node features. Once the number of node features is consistent across clients, it becomes the second category. A GNN model is trained with the graph, and GNN model weights are uploaded to the server for FL aggregation. 
However, the current solution limits the node feature choices to a few predefined network architectures, which is not efficient and flexible.

\section{Implementation}
\label{sec:Datasets}


Performance benchmarking is an essential factor for the long-term improvement of the FedGNNs research field. In this section, we review and discuss the applications with benchmarks, evaluation metrics, experiment evaluation designs, and platforms in existing FedGNNs literature.

\subsection{Applications}

There are several benchmark datasets developed for GNNs, including citation network datasets, social network datasets and chemical property datasets. FedGNNs test their algorithms on these datasets \cite{scardapane2020distributed,chen2021fedgraph,wang2021fl,zheng2021asfgnn,wang2020graphfl,chen2021fedgl,zhang2021subgraph,chen2022graph,ni2021vertical,ijcai2022p272,cheung2021fedsgc,xingbig,xie2021federated,zhu2021federated,pei2021decentralized,pei2021decentralized,he2021spreadgnn,hu2022fedgcn,zhu2021federated,xie2021federated,lin2020improving} with various data partition methods. 
FedGNNs also explore many GNNs applications in a decentralized setting with privacy concerns.
FedGNNs have been applied in knowledge graphs (KG) completion \cite{chen2021fede,zhang2022efficient,chen2022federated,peng2021differentially} and recommendation system tasks \cite{liu2021federated,wu2022federated,wu2021fedgnn} with privacy protection by considering one KG or one user as one client. Besides, income prediction \cite{lee2022privacy} and malicious transaction detection \cite{du2022malicious, peng2022domain} are also potential applications for FedGNNs.
FedGNNs have been used in FL applications with graph-structured information.
In computer vision applications,  
FedGNNs can improve the image classification performance \cite{caldarola2021cluster,lalitha2019peer,dinh2021new} by making close clients have similar local models.
It can also help the image classification training with heterogeneous network structures across clients by converting the local neural network structure into a graph \cite{litany2022federated}.
Healthcare applications \cite{bayram2021federated,lou2021stfl,peng2022fedni} can be addressed with FedGNNs when the data contain graph structures or there are relationships among patients.
For example, brain imaging data can be parcelled into different regions of interest (ROIs) with each ROI as one node in the graph. The population graph between patients can improve disease prediction with the help of FedGNNs \cite{bayram2021federated}. 
Transportation can leverage FedGNNs in many situations, such as traffic flow prediction \cite{meng2021cross,li2022federated,yuan2022fedstn,zhang2021fastgnn}, object position prediction \cite{jiang2022federated}, or indoor localization \cite{wu2022multi,chen2018deep}.
Sensors or surveillance cameras can be modeled as clients connected by a map graph. Alternatively, the objects detected by the cameras can form local graph.

\begin{table*}[t!]
\centering
\caption{Summary of applications, datasets and the corresponding evaluation metrics in FedGNNs.}
\resizebox*{1\linewidth}{!}{
\begin{tabular}{|ll|l|l|l|l|}
\hline
\multicolumn{2}{|l|}{ \textbf{Applications}} & \textbf{Datasets} & \textbf{References} & \textbf{Research Problems} & \textbf{Evaluation Metrics} \\ \hline
\multicolumn{2}{|l|}{Citation Network} & \makecell[l]{
Citeseer, Cora, PubMed, Cora-Full,\\ Physics, Coauthor CS, MSAcedemic, \\Amazon2M, ACM, AIFB, BGS,\\Wiki,  Aminer, ogbn-arxiv, DBLP} & 
\makecell[l]{\cite{scardapane2020distributed,chen2021fedgraph,wang2021fl},\\
\cite{zheng2021asfgnn,wang2020graphfl,chen2021fedgl},\\
\cite{zhang2021subgraph,chen2022graph,ni2021vertical},\\
\cite{ijcai2022p272,cheung2021fedsgc}}
& Node classification & Accuracy \\ \hline
\multicolumn{2}{|l|}{Social Network} & \makecell[l]{REDDIT, REDDIT-BINARY, COLLAB, \\ IMDB\_BINARY, IMDB\_MULTI,\\ GITHUB\_STARGAZERS,\\
Zachary karate club network\\
NEGAME} & 
\cite{xingbig,xie2021federated}
& Node classification & Accuracy \\ 
\hline
\multicolumn{2}{|l|}{\multirow{2}{*}{Chemical property prediction}} & \makecell[l]{FreeSolv,QM9,\\ Lipophilicity, ESOL}& 
\cite{zhu2021federated,pei2021decentralized}
& \makecell[l]{Graph embedding\\ regression} & MAE, MSE, RMSE \\ \cline{3-6}
&  & \makecell[l]{Tox21,SIDER,BACE, ClinTox, BBBP, \\ ENZYMES, D\&D, PROTEINS, PPIN, \\ MUTAG, NCI1} & 
\makecell[l]{\cite{pei2021decentralized,he2021spreadgnn,hu2022fedgcn},\\\cite{zhu2021federated,xie2021federated,lin2020improving}}
& Graph classification & \makecell[l]{Accuracy, F1-score, \\ROC-AUC} \\ \hline
\multicolumn{2}{|l|}{Knowledge graph completion} & \makecell[l]{FB15k-237, NELL-995, WN18RR, \\ DDB14, Dbpedia, Yago} & 
\makecell[l]{\cite{chen2021fede,zhang2022efficient,chen2022federated},\\
\cite{peng2021differentially}}
& Link prediction & MRR, Hits@N \\ \hline
\multicolumn{2}{|l|}{Recommendation} & \makecell[l]{Ciao, Epinions, Filmtrust, \\Flixser, Douban, YahooMusic, \\MovieLens(ML)-100K/1M/10M} &
\cite{liu2021federated,wu2022federated,wu2021fedgnn}
&\multirow{2}{*}{\makecell[l]{Node embedding \\regression}} & \multirow{2}{*}{MAE, MSE, RMSE}\\ \cline{1-4}
\multicolumn{2}{|l|}{Income Prediction} & UCI dataset (Adult) & 
\cite{lee2022privacy}
& &  \\ \hline
\multicolumn{2}{|l|}{Malicious Transaction detection} & Elliptic Data Set, Twi-Bot-20 & 
\cite{du2022malicious, peng2022domain}
& Node classification & F1-score \\ \hline
\multicolumn{1}{|l|}{Computer Vision} & Image classification & \makecell[l]{CelebA, MNIST, MedMNIST,\\
Federated Extended MNIST, \\CIFAR-10, CIFAR-100\\Chest X-ray, Vehicle sensor, \\Human Activity Recognition} & 
\makecell[l]{\cite{caldarola2021cluster,litany2022federated,lalitha2019peer,dinh2021new}}
&Classification & Accuracy\\ \hline
\multicolumn{1}{|l|}{\multirow{3}{*}{Healthcare}} & \makecell[l]{Brain template  \\ estimation} & ABIDE-I, OASIS-2  & 
\cite{bayram2021federated}
& Graph learning & \makecell[l]{Frobenius distance,\\ MAE} \\ \cline{2-6} 
\multicolumn{1}{|l|}{} & \makecell[l]{Sleeping stage \\classification} & ISRUC\_S3 & 
\cite{lou2021stfl}
&Classification & Accuracy, F1-score \\ \cline{2-6} 
\multicolumn{1}{|l|}{} & Disease prediction & ADNI, EHRs  &
\cite{peng2022fedni}
&Node classification & \makecell[l]{Accuracy, \\ROC-AUC} \\\hline
\multicolumn{1}{|l|}{\multirow{4}{*}{Transportation}} & 
\makecell[l]{Traffic flow \\ prediction} & \makecell[l]{PEMS-BAY, METR-LA, TaxtNYC, \\TaxtBJ, PeMSD4, PeMSD7,\\ Citi-Bike Dataset from New York City, \\ Citi-Bike Dataset from Washington DC, \\ Citi-Bike Dataset from Chicago} & 
\makecell[l]{\cite{meng2021cross,li2022federated,yuan2022fedstn},\\
\cite{zhang2021fastgnn}}
&\multirow{3}{*}{\makecell[l]{Node embeding\\ regression}} & \multirow{3}{*}{MAE, RMSE, MAPE} \\ \cline{2-4}
\multicolumn{1}{|l|}{} & \makecell[l]{Object position\\ prediction} & Stanford Drone Dataset (SDD) &
\cite{jiang2022federated}
& &  \\ \cline{2-4}
\multicolumn{1}{|l|}{} & Indoor localization & \makecell[l]{Received signal strenth (RSS) dataset \\of a shopping mall}  &
\cite{wu2022multi,chen2018deep}
& &  \\ \cline{2-6} 
\multicolumn{1}{|l|}{} & \makecell[l]{Airport busyness\\ prediction} & \makecell[l]{Brazilian air-traffic network \\European air-traffic network} & 
\cite{mei2019sgnn}
& Node classification & Accuracy\\ \hline
\end{tabular}
}
\label{tab:Application_Datasets}
\end{table*}

\subsection{Evaluation Metrics}
The usage of evaluation metrics depends on the learning task. In general, there are two types of tasks: 1) classification task and 2) regression task.
For classification tasks (e.g., node classification, graph classification, image classification), accuracy, F1-score and ROC-AUC have been adopted as the evaluation metrics. 
For regression tasks (e.g., node embedding regression, graph embedding regression, graph learning, link prediction), the following evaluation metrics are adopted:
\begin{itemize}
    \item For node and graph embedding regression tasks (e.g., recommendation system \cite{liu2021federated}, traffic flow prediction \cite{zhang2021fastgnn}), Mean Absolute Error (MAE), Mean Squared Error (MSE), Root Mean Squared Error (RMSE) and Mean Absolute Percentage Error (MAPE) are adopted as the evaluation metrics \cite{liu2021federated,zhu2021federated,jiang2022federated,qiu2022privacy} to measure the distance between predicted values and the ground-truth.

    \item For graph learning tasks (e.g., brain connectivity estimation \cite{bayram2021federated}), apart from the MAE, the graph learning performance can be measured by the Frobenius distance between the estimated graphs and the ground-truth.
    
    \item For link prediction tasks (e.g., knowledge graph completion \cite{chen2021fede,zhang2022efficient,chen2022federated, peng2021differentially}), the link prediction performance can be evaluated with Mean Rank, Mean Reciprocal Rank (MRR) and the proportion of correct entities in top $N$ ranked entities (Hits@N). 
    
\end{itemize}
FedGNNs applications, the corresponding datasets, research problems and evaluation metrics are summarized in Table \ref{tab:Application_Datasets}. 




\subsection{FedGNNs Experimental Evaluation Design}
\label{ssec:DataPartition}

Since there are few real cross-silo graph datasets, the majority of works simulate the distributed setting by performing partitioning on the public datasets summarized in Table \ref{tab:Application_Datasets}.
Here, we discuss the experimental evaluation design in the FedGNNs literature according to the different scenarios in the main taxonomy. 

In the FL-Assisted GNNs setting, an inter-client graph exists between clients with various local data types, such as graph data, image data or temporal data, etc. 
For the local data partition, clients with IID and non-IID data distribution have different data partition methods.
Besides, there are several ways to build an inter-client graph. 
These methods are summarized as follows. 

To construct FL clients with the IID data distribution, samples are distributed evenly and randomly to clients. A sample can be an image \cite{lalitha2019peer,litany2022federated} (e.g., an image from MNIST), a time-sequence data from a sensor \cite{wu2022multi} or a protein graph from some bio-medicine dataset \cite{tao2022semigraphfl}.
A sample can also be a node in the graph \cite{scardapane2020distributed}. However, the partition method is different from the above. 
Each client selects some seed nodes randomly and expands the graph with a breadth-first search on the original entire graph to get their local data.

To construct FL clients with non-IID data distributions, there are five main approaches:
\begin{enumerate}
    \item \textbf{Imbalanced partition}: Different clients have different numbers of samples. It can be achieved by distributing the samples using a Latent Dirichlet Allocation (LDA) \cite{pei2021decentralized,he2021spreadgnn}.
    \item \textbf{Clustering partition}: Different clients have different clusters. It can be achieved by dividing the samples into clusters using a clustering algorithm (e.g. k-means) \cite{wu2022multi,tao2022semigraphfl}. 
    \item \textbf{Label distribution skew partition}: Different clients have different subsets of label classes. 
    It can be achieved by making one client maintain data only from one class or a subset of classes \cite{caldarola2021cluster,lee2022privacy}.
    \item \textbf{Natural identity partition}: Different clients have different characteristics. For example, in the traffic flow dataset, one sensor is assigned to one client \cite{meng2021cross}. 
    In human activity recognition, one person with his/her data is considered as one client \cite{dinh2021new}. In Nature Language Processing (NLP) datasets, documents from one domain are assigned to one client \cite{xingbig}. 
    \item \textbf{Synthetic data generation}: Generate synthetic data points for clients following Gaussian distribution with different variances \cite{rizk2021graph,gogineni2022decentralized,Lu2020decentralized}.
\end{enumerate}
The inter-client graphs can be obtained as follows: 
\begin{enumerate}
    \item \textbf{Natural graph}: Some datasets contain a natural graph topology that can be used as an inter-client graph directly. 
    For example, in traffic flow data, the road map can be used as an inter-client graph with each sensor as one client \cite{meng2021cross}. 
    In the NLP dataset application, the syntactic structure can also work as the graph between different domains as clients \cite{xingbig}.
    And the wireless communication network can also serve as it between routers \cite{li2022power}.
    \item \textbf{Simulated graph based on clients' similarities}: There is no natural graph topology in some datasets. But the inter-client graph can be calculated based on the clients' information, such as clients' embedding \cite{wu2022multi,caldarola2021cluster,lee2022privacy,lee2022privacy,dinh2021new}.
    \item \textbf{Synthetic graph}: A fully connected graph \cite{tao2022semigraphfl,scardapane2020distributed}, ring connected graph \cite{lalitha2019peer}, or just a random generated graph can be used as the inter-client graph \cite{rizk2021graph,gogineni2022decentralized,Lu2020decentralized}. 
\end{enumerate}

In the GNN-Assisted FL setting, clients maintain a set of graphs or a set of nodes (one graph). We summarize the main data partition methods for both horizontal and vertical FedGNNs below.

In the horizontal FedGNNs scenario, clients have different node IDs. 
If each client has a set of graphs, the data partition methods for IID and non-IID are very similar to the FL-Assisted GNNs setting. 
For IID, graph samples are distributed to clients evenly and randomly \cite{hu2022fedgcn,gurler2022federated,bayram2021federated,balik2022investigating,jiang2022federated,peng2022fedni}.
For non-IID, imbalanced partition \cite{zhu2021federated}, label distribution skew partition (graphs of one class are assigned to one client) \cite{lou2021stfl} and natural identity partition (one dataset is considered as one client) \cite{xie2021federated} are applied.

If each client has only one graph, it is more complex for both settings to partition one entire graph into several sub-graphs. 
For IID, two methods are applied: 1) non-overlapping partition by randomly dividing the entire graph into several sub-graphs \cite{zhang2021fastgnn} and 2) overlapping partition by randomly assigning a portion of nodes or edges to one client\cite{chen2021fede,zhang2022efficient, chen2022federated,chen2021fedgl,chen2021fedgraph,wang2020graphfl}.  
For non-IID, strategies used in FL-Assisted GNNs settings still work albeit with some changes. 
For clustering partition, nodes, instead of general samples, are clustered into groups \cite{wang2021fl,zhang2021subgraph,zhang2022graph}.
For label distribution skew partition, each client selects most nodes from major classes and few nodes from minor classes \cite{zhang2022fedego,zheng2021asfgnn,du2022federated}. 
For natural identity partition, more identity units are applied to partition the data.
In the recommendation application, one user with its interactions is considered as one client \cite{liu2021federated,wu2022federated,wu2021fedgnn}.
In Knowledge Graph (KG) completion application, one KG is considered as one client \cite{peng2021differentially}. 
In some citation network applications, papers published in the same year are assigned to one client \cite{liu2022federated,du2022federated}. 

In the vertical FedGNNs scenario, clients have the same node IDs. 
For the incomplete graph setting, 
if there are three clients, then each of them gets one of
the node features, graph topology, and node labels of graph data \cite{mei2019sgnn}. If there are only two clients, then one will get one component of graph data and the other gets the rest \cite{cheung2021fedsgc}.
For the complete graph setting, node features are divided evenly to clients with the graph topology retained by all clients \cite{chen2022graph,ni2021vertical,ijcai2022p272,li2022federated}.

\subsection{FedGNNs Platforms}
Currently, there are two FedGNNs platforms. 
FedGraphNN\cite{he2021fedgraphnn} is an open-source platform supporting 3 GNN models and 2 FL aggregation methods. It has collected 36 graph datasets and partitioned them into distributed silos, forming a promising FedGNNs benchmarking tool.
FederatedScope-GNN \cite{Wang2022FederatedScopeGNN} consists of an event-driven FedGNNs framework with two components: 1) ModelZoo and 2) DataZoo. ModelZoo provides comprehensive GNN models (e.g., GCN \cite{kipf2016semi}, GAT \cite{velickovic2018graph}, GraphSage \cite{hamilton2017inductive}) used in the clients, and some of the existing FedGNNs models (e.g., FedSage+, FedGNN, GCFL+). DataZoo provides a collection of splitting strategies for distributing a given graph dataset among FL clients.

\section{Promising Future Research Directions}
\label{sec:Future_Direction}
As an emerging field, FedGNNs research is starting to gain traction. Nevertheless, in order for this technology to effectively deal with challenges in real-world applications, many problems remain to be addressed. Here, we highlight seven of them which hold promising opportunities: 
\begin{enumerate}
    \item \textbf{Robust FedGNNs against malicious attacks.}
    By sharing node embeddings, graph topology and model parameters, FedGNNs have large attack surfaces. Although some works attempt to address this issue by leveraging differential privacy \cite{peng2021differentially,zhang2021fastgnn,wu2021fedgnn} or cryptographic methods \cite{chen2021fedgraph,ijcai2022p272,rizk2021graph,zheng2021asfgnn}, they are designed to guard against only semi-honest attackers. Additional research is needed to explore how FedGNNs can be made more robust in the face of malicious privacy attacks.

    \item \textbf{Explainable FedGNNs to improve interpretability.}
    Works on the explainability of GNNs \cite{yuan2020explainability} and FL \cite{li2021survey} are starting to emerge. FedGNN involves complex model structures and training processes. Thus, achieving explainability \cite{Zhang-Yu:2022IJCS} under this setting is even more challenging. The incorporation of explainability into FedGNNs needs to jointly consider the needs for interpretability by the stakeholders involved while balancing the goals of preserving privacy and training models efficiently. 
    
    \item \textbf{Efficient FedGNNs for large-scale graph data.}
    Existing FedGNNs are generally studied with small-scale distributed datasets. Thus, communication efficiency has not yet been adequately considered. 
    However, in order to scale FedGNNs up to large-scale graph data (e.g., knowledge graphs), communication overhead can be an important bottleneck since the clients often adopt multi-layer GNN models with a large number of model parameters to be transmitted.

    \item \textbf{Fair FedGNNs for clients with biased graphs.} Graph data consist of the graph topology and data. Existing FedGNNs mainly focus on the non-IID problem in data distribution, while ignoring the non-IID problem in graph topology distribution across FL clients. Different clients may own graphs with different properties. For example, some clients have graphs from one class with high node degrees, while others having graphs low node degrees \cite{chen2022ba,dong2022structural}. Such biased graphs can affect outcomes, or even cause harm in the FedGNNs \cite{Zhang-et-al:2022IJCS}. 
    Thus, achieving fairness in this setting is important. 
    
    \item \textbf{Continual FedGNNs training for new clients.}
    In GNN-Assisted FL, clients are treated as nodes in the graph. As new clients join, a well-trained global model can be applied to them directly. However, in the decentralized setting, since there is no server to coordinate the global training, all clients need to re-train their local models when new clients join. Thus, to improve the training efficiency, it is necessary to develop continual learning algorithms for decentralized FedGNNs.
    
    \item \textbf{Comprehensive FedGNNs frameworks supporting diverse GNN models.} Various GNN algorithms have been rapidly emerging in recent years. However, existing FedGNNs only employ a limited set of GNN models (e.g., GCN, GAT, GraphSage).
    Thus, to make full use of GNNs to assist FedGNNs in solving more difficult problems, comprehensive FedGNNs frameworks with more GNNs algorithms or strategies are required.


    \item \textbf{Realistic cross-silo graph datasets for benchmarking.}
    Existing FedGNNs are mostly evaluated with graph data partitioned artificially. Nevertheless, the long-term development of this field still requires realistic and large-scale federated graph datasets to be made available to support experimental evaluations under settings close to practical applications. Real-world graph datasets, such as healthcare datasets, recommender systems and knowledge graphs, can be useful starting points.

\end{enumerate}

\section*{Acknowledgments}
This research/project is supported by the National Research Foundation, Singapore and DSO National Laboratories under the AI Singapore Programme (AISG Award No: AISG2-RP-2020-019); Alibaba Group through Alibaba Innovative Research (AIR) Program and Alibaba-NTU Singapore Joint Research Institute (JRI) (Alibaba-NTU-AIR2019B1), Nanyang Technological University, Singapore; the RIE 2020 Advanced Manufacturing and Engineering (AME) Programmatic Fund (No. A20G8b0102), Singapore; Nanyang Technological University, Nanyang Assistant Professorship (NAP); and Future Communications Research \& Development Programme (FCP-NTU-RG-2021-014).

\bibliography{ref}
\bibliographystyle{IEEEtran}


\begin{IEEEbiography}[{\includegraphics[width=1in,clip,keepaspectratio]{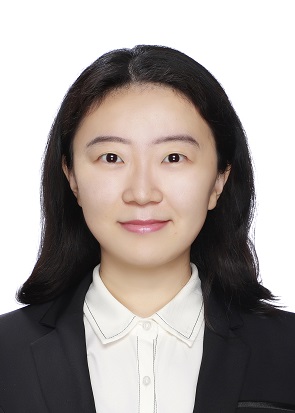}}]{Rui Liu} received the BEng degree from the Harbin Institute of Technology (HIT), Harbin, China, in 2014 and the Ph.D. degree from the Singapore University of Technology and Design (SUTD), Singapore, in 2019. Currently, she is a research fellow at the School of Computer Science and Engineering (SCSE), Nanyang Technological University (NTU), Singapore. Her research focuses on graph neural networks, federated learning, and brain-computer interfaces. 
\end{IEEEbiography}

\begin{IEEEbiography}[{\includegraphics[width=1in,clip,keepaspectratio]{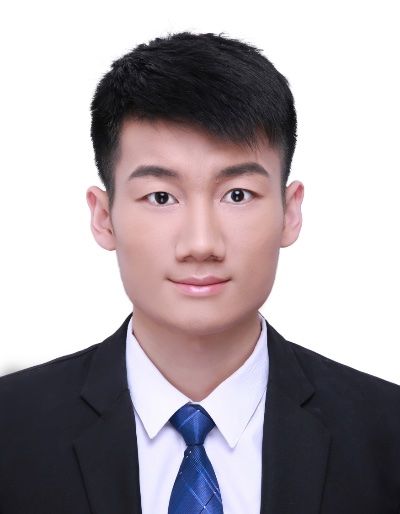}}]{Pengwei Xing} received his Bachelor’s and Master's degrees in computer science from Henan University and Tianjin University, China in 2016 and 2019, respectively. He is currently working toward his PhD in the School of Computer Science and Engineering (SCSE), Nanyang Technological University (NTU), Singapore. His research mainly focuses on federated learning and graph learning.
\end{IEEEbiography}

\begin{IEEEbiography}[{\includegraphics[width=1in,clip,keepaspectratio]{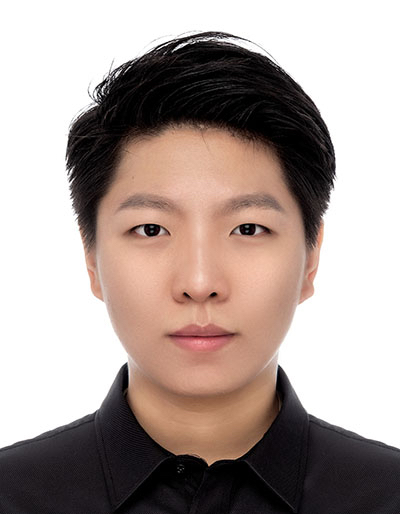}}]{Zichao Deng} received the BEng degree from the Nanyang Technological University (NTU), Singapore, in 2018. Currently, he is a Ph.D. student at the School of Computer Science and Engineering (SCSE), NTU. His research focuses on federated graph learning. 
\end{IEEEbiography}

\begin{IEEEbiography}[{\includegraphics[width=1in,height=1.28in,clip,keepaspectratio]{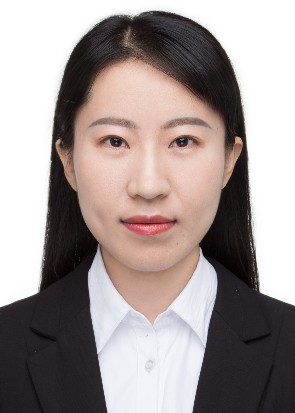}}]
	{Anran Li} received her BS degree from Anhui University of Science and Technology, China, in 2016, and the Ph.D. degree from University of Science and Technology of China, China, in 2021. 
 She is currently a research fellow in the School of Computer Science and Engineering, Nanyang Technological University, Singapore. 
	Her research interests mainly focus on data quality assessment, federated learning and mobile computing.
\end{IEEEbiography}

\begin{IEEEbiography}[{\includegraphics[width=1in,clip,keepaspectratio]{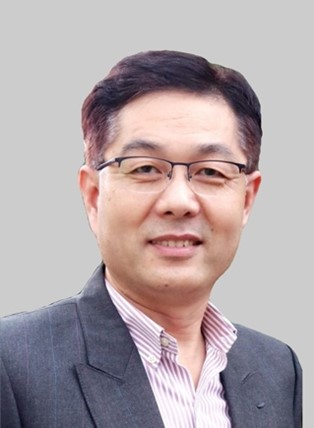}}]{Cuntai Guan} (Fellow IEEE) is currently a President’s Chair Professor in the School of Computer Science and Engineering, Director of Artificial Intelligence Research Institute, Director of Centre for Brain-Computing Research, and Co-Director of S-Lab for Advanced Intelligence at the Nanyang Technological University, Singapore. His research interests include brain-computer interfaces, machine learning, neural signal and image processing, neural and cognitive rehabilitation, and artificial intelligence. He is a recipient of the Annual BCI Research Award, the IES Prestigious Engineering Achievement Award, the Achiever of the Year (Research) Award, King Salman International Award for Disability Research, and the Finalist of President Technology Award. He is also a Fellow of AIMBE, Fellow of NAI, and Fellow of the Academy of Engineering Singapore.  

\end{IEEEbiography}

\begin{IEEEbiography}[{\includegraphics[width=1in,clip,keepaspectratio]{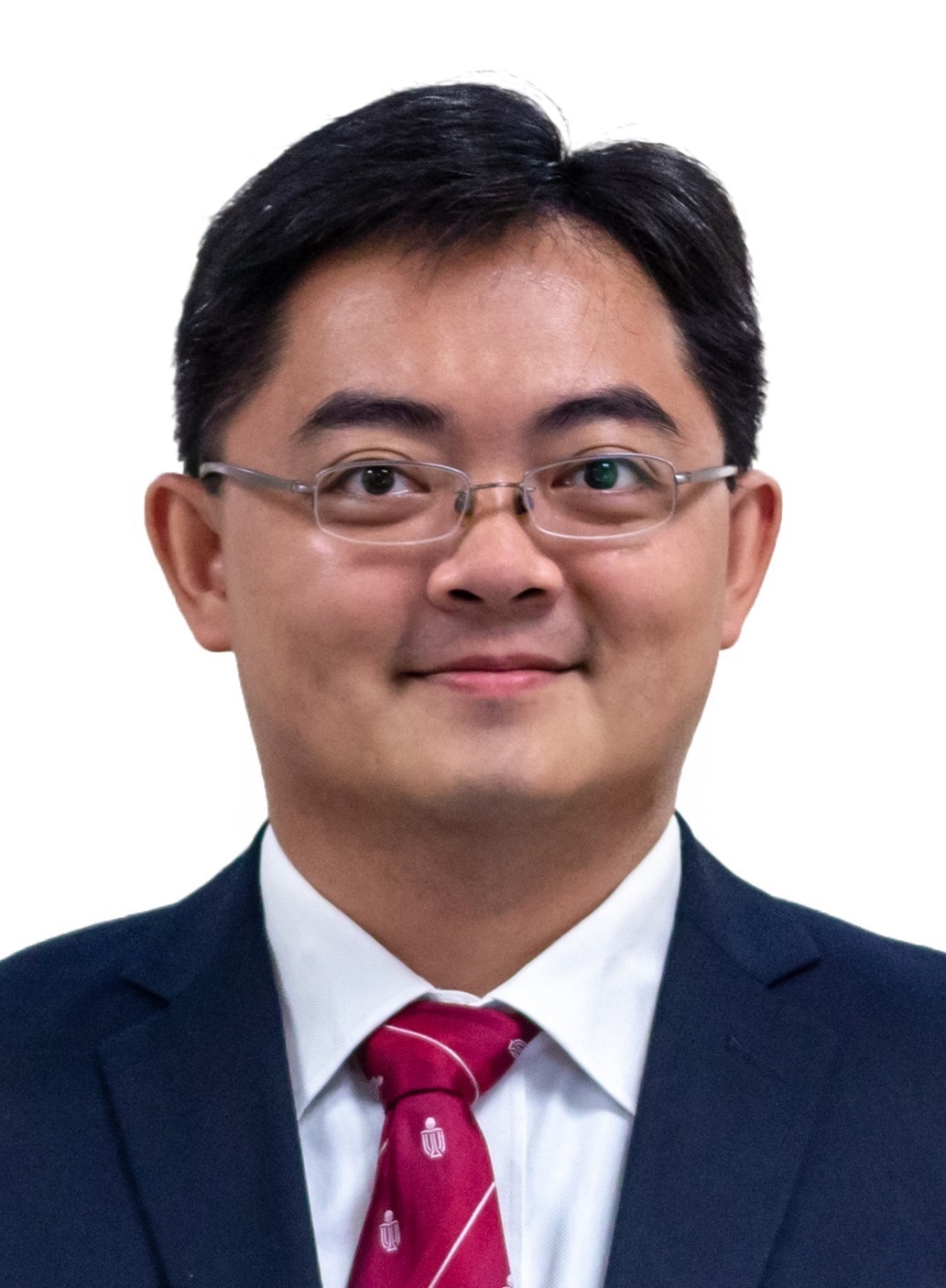}}]{Han Yu} is a Nanyang Assistant Professor (NAP) in the School of Computer Science and Engineering (SCSE), Nanyang Technological University (NTU), Singapore. He held the prestigious Lee Kuan Yew Post-Doctoral Fellowship (LKY PDF) from 2015 to 2018. He obtained his PhD from the School of Computer Science and Engineering, NTU. His research focuses on federated learning and algorithmic fairness. He has published over 200 research papers and book chapters in leading international conferences and journals. He is a co-author of the book \textit{Federated Learning} - the first monograph on the topic of federated learning. His research works have won multiple awards from conferences and journals. He is a \textit{Senior Member} of CCF and IEEE.
\end{IEEEbiography}

 




\vfill

\end{document}